\documentstyle[12pt]{article}
\renewcommand{\baselinestretch}{1.3}
\newcommand{\be}{\begin{equation}}
\newcommand{\ee}{\end{equation}}

\begin{document}
\title{Spectrum of atomic radiation at sudden perturbation}
\author{Victor I. Matveev\\
Heat Physics Department of Uzbek Academy of Sciences,\\
28 Katartal St., 700135 Tashkent, Uzbekistan}

\begin{titlepage}
\maketitle

\begin{abstract}
A general expression for the spectrum of photons emitted by
atom  at sudden perturbation is obtained.  Some concrete examples
of application of the obtained result are considered.
The conclusion about the coherence of radiation of the atomic electrons
under the such  influences is made. \\
PACS number: 32.30.*
\end{abstract}
\end{titlepage}

\normalsize

  It is  known many examples when the excitation or ionization of atoms
occurs as result of
the action of sudden perturbations.
First of all these are atomic excitation or ionization
in the  nuclear reactions [1,2]. For example in  $ \beta $-decay
of nucleus, when the fast $\beta$-electron's escape is perceived
by atomic electrons as a sudden changing  of nuclear charge
or in neutron impact with nucleus, when the sudden
of momentum transfer  to the nucleus occurs etc.
The sudden approximation [3] can be used for
consideration multielectron transition in complex atoms,
when transition occurring in internal shells,
are perceived by relatively slow electrons of external shells as
instantaneous (see [4,5]).
As a result of action of sudden perturbation can be considered
inelastic processes
in the  collisions of fast multicharged ions with atoms [6 - 12] and in the
collisions of charged particles with highly-excited atoms [13].
After action of sudden perturbation, the excited atom
can  relax with
radiation of
photons belonging to known spectrum of isolated atom.
However, if sudden perturbation causes the change of velocities of atomic
electrons, atom can radiate during the action of perturbation.
Classical analogue of such a problem
is the  [14]  radiation of a free electron under the
sudden changing of velocity.
In many practically important cases  perturbation
is not sufficiently small to use a  perturbation theory. However
the situations when the time of action of perturbation
is considerably less than the characteristic atomic time
that enables one to solve the
problem without restricting the value of perturbation
(see for instance [9,15-17]).

Thus, it is necessary to state a general problem on the spectrum of
photons  emitted by atom during the time of action of sudden perturbation,
i.e. - on the spectrum of photons emitted simultaneously by all atomic
electrons as a result of action of perturbation.
In this paper we derive a general expression for the spectrum of photons
emitted by the atom under sudden perturbation and apply this result to some
concrete processes.

 Consider "collision"  type sudden perturbation [3],
when the perturbation $ V(t) \equiv V({\bf r}_{a},t)\; $  (where
 ${\bf r}_{a}$ are the coordinates of atomic electrons)
acts only during the time $ \tau $, which is much smaller
than the characteristic period of
unperturbed atom, describing by Hamiltonian $H_0$.
To be definite  we will assume that $V(t)$ is not equal zero near $t=0$ only.
Then in the exact solution of Schr\"odinger equation (
atomic units are used throughout in this paper)
$$
i \frac{\partial \psi}{\partial t}=(H_0+V(t))\psi
$$
one can neglect by evolution of  $ \psi $ (during the time $ \tau $) caused
by
unperturbed Hamiltonian $H_0$.
Therefore the transition amplitude of
atom from the initial state $ \varphi _{0} $ to a final state
$ \varphi _{n} $, as a result of actions of sudden perturbation
 $V(t)$, has  the form [3,6]:
\be
a_{0n}=\langle \varphi _{n}\mid
exp(-i\int\limits_{-\infty}^{+\infty} V(t)dt)\mid \varphi _{0}\rangle \;,
\ee
where $ \varphi _{0} $ and $ \varphi _{n} $ belong to the full
set of orthonormalized eigenfunctions of the unperturbated Hamiltonian $H_0$,
i.e. $ H_0\varphi _{n}=\epsilon _{n}\varphi _{n} $.

Thus in the sudden perturbation  approximation the evolution of the
initial state has the form
\be
\psi _{0}(t)=exp(-i\int\limits_{-\infty}^{t}V(t')dt')\varphi _{0}\;,
\ee
where $ \psi _{0}(t) $ satisfies the equation
\be
i\frac{\partial \psi _{0}(t)}{\partial t}=V(t)\psi _{0}(t)\;,
\ee
and $ \psi _{0}(t)\rightarrow \varphi _{0} $ under $ t \rightarrow
-\infty $.
Let's introduce full and orthonormal set of functions
\be
\Phi _{n}(t)=exp(i\int\limits_{t}^{+\infty}V(t' )dt')\varphi _{n}\; ,
\ee
obeying eq. (3), and $ \Phi _{n}(t) \rightarrow \varphi
_{n} $ at $ t \rightarrow +\infty $.
Obviously the  amplitude (1) can be rewritten as
$$
a _ {0n} = \langle \Phi _ {n} (t) \mid \psi _ {0} (t) \rangle.
$$

Therefore the  one photon radiation amplitude can be calculated in the 
first order of perturbation theory (as a corrections to the states
(2) and (4))  over the interaction of atomic electrons with electromagnetic
field [18,19]
$$
W = -\sum \limits_{a,{\bf k},\sigma } \biggl(\frac{2\pi }{\omega
  }\biggr) ^{\frac{1}{2}}{\bf u} _{{\bf k}\sigma}(a_{{\bf k}\sigma}^{+}
 e^{-i{\bf k}{\bf r}_{a}}+a_{{\bf k}\sigma} e^{-i{\bf k}{\bf r}_{a}})
{\bf {\hat p}}_{a}\;\;,
$$
where $ a_{{\bf k}\sigma}^{+} $  and $ a_{{\bf k}\sigma} $ are the
creation and annihilation operators of the photon with a frequency
$ \omega $, momentum {\bf k} and polarization
 $\sigma,\; (\sigma =1,2),  {\bf u}_{{\bf k}\sigma } $  are the unit
vectors of polarization,  $ {\bf r}_{a} $ are the coordinates of atomic
electrons ($ a =1,..,Z_{a}$),
here $ Z_{a} $  is the number of atomic electrons, ${\bf {\hat p}}_{a} $
are the momentum  operators of atomic electrons.
Then in the dipole approximation the amplitude of emission of photon with
simultaneous transition of atom
from the state $
\varphi _{0} $ to a state $ \varphi _{n} $ has the form
$$
b_{0n}(\omega ) = i\biggl(\frac{2\pi}{\omega}\biggr) ^{\frac{1}{2}}
{\bf u}_{{\bf k}\sigma}\int\limits_{-\infty}^{+\infty}dt e^{i\omega t}
\langle \Phi
 _{n}(t)\mid \sum\limits_{a}{\bf {\hat p}}_{a} \mid \psi _{0}(t) \rangle .
$$
Integrating this expression by parts over the time and omitting the terms
vanishing
 (at $t\rightarrow \pm \infty $) in turning off the interaction with
electromagnetic field we have
\begin{eqnarray}
\displaystyle
b_{0n}(\omega ) = i\biggl(\frac{2\pi}{\omega
}\biggr) ^{\frac{1}{2}}{\bf u}_{{\bf k}\sigma}
\int\limits_{-\infty}^{+\infty} dt
\frac{e^{i \omega t}}{\omega}
\times
\nonumber \\  \times
\langle \varphi _{n} \mid
 \sum\limits_{a} \frac{\partial V(t)}{\partial {\bf r}_{a}} exp \bigl(-i
  \int\limits_{-\infty}^{+\infty} V(t^{'})dt^{'} \bigr) \mid \varphi _{0}
   \rangle .
\end{eqnarray}
Summing  $ \mid b_{0n}(\omega)\mid ^{2} $
 over polarization and integrating over the photon's emission angles
and summing, after this, over all final states of the atom
 $ \varphi _{n} $,
we find the total radiation spectrum
\be
\frac{dW}{d\omega}= \frac{2}{3\pi} \frac{1}{c^{3}\omega} \langle
 \varphi _{0} \mid \sum\limits_{a} \frac{\partial
\tilde V^{*}(\omega)}{\partial {\bf r}_{a}} \sum\limits_{b} \frac{\partial
\tilde V(\omega)}{\partial {\bf r}_{b}} \mid \varphi _{0} \rangle ,
\ee
where c = 137  a.u. is the speed of light,
\be
\tilde V(\omega)= \int\limits_{-\infty}^{+\infty} V(t) e^{i \omega t} dt.
\ee
Thus we have obtained the radiation spectrum of atom during the time of
sudden perturbation $V(t)$.

As an application we consider the radiation spectrum of atom in the sudden
transmission of momentum $ \bf p $ to the atomic electrons when
 $V(t)$ has the (widely used for collision problems) form
\be
V(t) = {\bf f} (t) \sum\limits_{a} {\bf r}_{a},\;\;{\bf p} =
 \int\limits_{-\infty}^{+\infty} dt {\bf f}(t) ,
\ee
and ${\bf f}(t)$ is the perturbing force which not depends on ${\bf r}_{a}$
and interacts during a time $ \tau $ that is considerable less than the
characteristic periods of the unperturbed atom.
The total radiation spectrum (6) in this case has the form
\be
\frac{dW}{d\omega } = \frac{2}{3\pi} \frac{1}{c^{3} \omega } \mid
 {\bf \tilde f} (\omega) \mid ^{2} \cdot Z_{a}^{2} ,
\ee
where  ${\bf \tilde f}(\omega), $ is the  Fourier transform of the functions
 ${\bf f}(t)$, defined according to (7),
$Z_{a}$ is the number of atomic electrons.
In this case the spectrum coincides (after producting to $ \omega $)
with the radiation spectrum of the classical particle with mass equal to
electron's one
and with charge $ Z_{a}$, moving in the field of homogeneous forces
${\bf f}(t)$.
This gives us the information about the value of the spectrum (9).
Since ${\bf f}(t)$ $ \not= 0 $ just during the time $ \tau $ ,
and the spectrum (9) is proportional to
 $ \mid {\bf \tilde f}(\omega) \mid ^{2} $, only the photons belonging to
continuum with characteristic frequencies
 $ \omega \leq 1/ \tau $ can be emitted by atom.

Analogously one can consider the radiation of atom in the "switching" type
 sudden perturbation (we use the classification of sudden perturbations
introduced in [3]).

Formula (5) allows one to obtain the spectrum of photons in the transition
of atom
 from the state $\varphi _{0} $ to a state $ \varphi _{n} $ under the
influence of
perturbation (8):
\be
\frac{dw_{0n}}{d \omega } = \frac{2}{3\pi} \frac{1}{c^{3} \omega }
 \mid {\bf \tilde f}(\omega) \mid ^{2} Z_{a}^{2} \mid \langle \varphi _{n}
\mid exp(-i
{\bf p} \sum\limits_{a} {\bf r}_{a}) \mid \varphi _{0} \rangle
\mid ^{2}.
\ee
Here  $dW/d \omega =\sum_{n} dw_{0n}/d \omega $, where
$\sum_{n}$ means summing over the complete set of atomic states.
Formula (10) allows one to express the relative contribution of transitions
with excitation to an arbitrary state $ \varphi _{n}$
to the total spectrum (9)
$$
\frac{dw_{0n}/d \omega }{dW/d \omega } =
\mid \langle \varphi _{n} \mid exp(-i {\bf p}
\sum\limits_{a} {\bf r}_{a}) \mid \varphi _{0} \rangle
\mid ^{2}
$$
via the well known [2] inelastic atomic formfactors
$\langle \varphi _{n} \mid exp(-i {\bf p}
\sum\limits_{a} {\bf r}_{a}) \mid \varphi _{0} \rangle$.

In the most simple case of instantaneous transferring to atomic electrons
the momentum
$ {\bf p} $, when in (8) ${\bf f}(t)=  {\bf p}  \cdot \delta (t), $
where $ \delta (t)$ is the Dirac $\delta $-function, then $
{\bf \tilde f}(\omega) = {\bf p} $ and spectrum (9) coincides,
after producting
 to $\omega $, with the radiation spectrum of free classical particle [14]
with charge
 $ Z_{a} $, which takes (suddenly) a velocity $ {\bf p}.$

As an another example we give the radiation spectrum in the influence of
pulse
having the Gausian form
$$
{\bf f}(t)={\bf f}_0 exp(-\alpha^2t^2)cos(\omega_0t)\;,
$$
respectively
$$
\tilde V(\omega)=
\frac{\sqrt{\pi}}{2\alpha}{\bf f_0} \sum\limits_{a}{\bf r}_{a}
\left\{exp\left[-\frac{(\omega-\omega_0)^2}{4\alpha^2}\right]
+exp\left[-\frac{(\omega+\omega_0)^2}{4\alpha^2}\right] \right\}\;.
$$
Therefore the radiation spectrum has the form
$$
\frac{dW}{d\Omega}=\frac{f_0^2}{6\Omega c^3 \alpha^2}
\left\{exp\left[-(\Omega+\Omega_0)^2\right]
+exp\left[-(\Omega-\Omega_0)^2\right]\right\}Z_a^2 \;,
\nonumber
$$
where for the sake of convenience the frequencies $\Omega=
\omega /(2\alpha)$ and
$\Omega_0=\omega_0 /(2\alpha)$ are introduced.

One should note an important generality of radiation at sudden perturbation,
namely, the radiation intensity for the multielectron atoms is proportional
to the square of the number of atomic electrons.
This fact allows one to conclude on the coherence of radiation of atomic
electrons under such type influences.

{\bf References} \\
1. A.B. Migdal, {\it Qualitative Methods in Quantum Theory}
  (Moscow: Nauka, 1975)\\
2. L.D. Landau and E.M. Lifshitz, {\it Quantum Mechanics}
   (Moscow: Nauka, 1989)\\
3. A.M. Dykhne, G.L. Yudin, Usp. Fiz. Nauk, {\bf 125}, 377 (1978).
   [Sov.Phys. Usp. {\bf 21}, 549 (1978)].\\
4. T. Aberg, in "{\it Photoionization and Other Probes of Many Electron
   Interactions}" (F. Wuillemier, ed. Plenum, New York, 1976, p. 49).\\
5. V.I. Matveev, E.S. Parilis, Usp. Fiz. Nauk, {\bf 138}, 573 (1982).
   [Sov. Phys.    Usp. 1982, {\bf 25}, 881 (1982)].\\
6. J. Eichler, Phys.Rev.A. {\bf 15}, 1856(1977).\\
7. G.L. Yudin, Zh.Eksp.Teor. Fiz. 1981, {\bf 80}, 1026 (1981).\\
8. J.H. McGuire, Advances in Atomic, Molecular and Optical
   Physics, {\bf 29}, 217 (1992).\\
9. V.I. Matveev, Phys.Part. Nuclei, {\bf 26}, 329 (1995).\\
10. P.K. Khabibullaev, V.I. Matveev, D.U. Matrasulov,
   J. Phys. B, {\bf 31}, L607 (1998).\\
11. V.I.Matveev, Kh.Yu.Rakhimov, D.U.Matrasulov.
   J.Phys. B, {\bf 32}, 3849 (1999).\\
12. V.I. Matveev, J. Phys. B, {\bf 24}, p. 3589 (1991).\\
13. I.C. Percival, in "{\it Atoms in Astrophysics}", (Edited by
    P.G. Burke, W.B. Eissner, D.G. Hammer and I.C. Percival
    Plenum Press, New York and London, 1983, p. 87-113.)\\
14. L.D. Landau, E.M. Lifshitz, The Classical Theory of Field
    (Moscow: Nauka, 1988).\\
15. M.Ya. Amusia, {\it The Bremsstrahlung}, (Moscow: Energoatomizdat, 1990.)\\
16. A.J. Baltz, Phys. Rev. A, {\bf 52}, 4970 (1995).\\
17. A.J. Baltz, Phys. Rev. Lett. {\bf 78}, p.1231 (1997).\\
18. The sudden perturbation $V(t)$  accounted in the functions
    $\Phi _{n}(t)$ ¨ $\psi _{0}(t)$  without limitation of value $V(t)$.\\
19. V.B. Berestetskii, E.M. Lifshitz and L.P. Pitaevskii,
    {\it Quantum Electrodynamics} (Moscow: Nauka, 1989)

\end{document}